\documentclass[prb,amssymb,reprint,superscriptaddress]{revtex4-1}

\usepackage{graphics}
\usepackage{mathtools}
\usepackage{float}
\usepackage{amsmath}

\begin{document}
\title{Signatures of Electronic Nematicity in (111) LaAlO$_3$/SrTiO$_3$ Interfaces}

\author{S. Davis}
\email[]{samueldavis2016@u.northwestern.edu}

\affiliation{Graduate Program in Applied Physics and Department of Physics and Astronomy, Northwestern University, 2145 Sheridan Road, Evanston, IL 60208, USA}
\author{Z. Huang}
\author{K. Han}

\affiliation{NUSNNI-Nanocore, National University of Singapore 117411, Singapore}  
\affiliation{Department of Physics, National University of Singapore 117551, Singapore } 
\author{Ariando}
\affiliation{NUSNNI-Nanocore, National University of Singapore 117411, Singapore}  
\affiliation{Department of Physics, National University of Singapore 117551, Singapore } 
\affiliation{NUS Graduate School for Integrative Sciences \& Engineering, National University of Singapore 117456, Singapore}

\author{T. Venkatesan}
\affiliation{NUSNNI-Nanocore, National University of Singapore 117411, Singapore}  
\affiliation{Department of Physics, National University of Singapore 117551, Singapore } 
\affiliation{NUS Graduate School for Integrative Sciences \& Engineering, National University of Singapore 117456, Singapore}
\affiliation{Department of Electrical and Computer Engineering, National University of Singapore 117576, Singapore}\affiliation{Department of Material Science and Engineering, National University of Singapore 117575, Singapore}

\author{V. Chandrasekhar}
\email[]{v-chandrasekhar@northwestern.edu}
\affiliation{Graduate Program in Applied Physics and Department of Physics and Astronomy, Northwestern University, 2145 Sheridan Road, Evanston, IL 60208, USA}

\date{\today}%

\maketitle
\textbf{Symmetry breaking is a fundamental concept in condensed matter physics whose presence often heralds new phases of matter.  For instance, the breaking of time reversal symmetry is traditionally linked to magnetic phases in a material, while the breaking of gauge symmetry can lead to superfluidity/superconductivity.  Nematic phases are phases in which rotational symmetry is broken while maintaining translational symmetry, and are traditionally associated with liquid crystals.  Electronic nematic states where the orthogonal in-plane crystal directions have different electronic properties have garnered a great deal of attention after their discovery in Sr$_3$Ru$_2$O$_7$,\cite{Borzi} multiple iron based superconductors,\cite{kim,kasa} and in the superconducting state of CuBiSe.\cite{naga,deg}  Here we demonstrate the existence of an electronic nematic phase in the two-dimensional carrier gas that forms at the (111) LaAlO$_3$ (LAO)/SrTiO$_3$ (STO) interface that onsets at low temperatures, and is tunable by an electric field.}

\begin{figure}[!]
\center{\includegraphics[width=7cm]{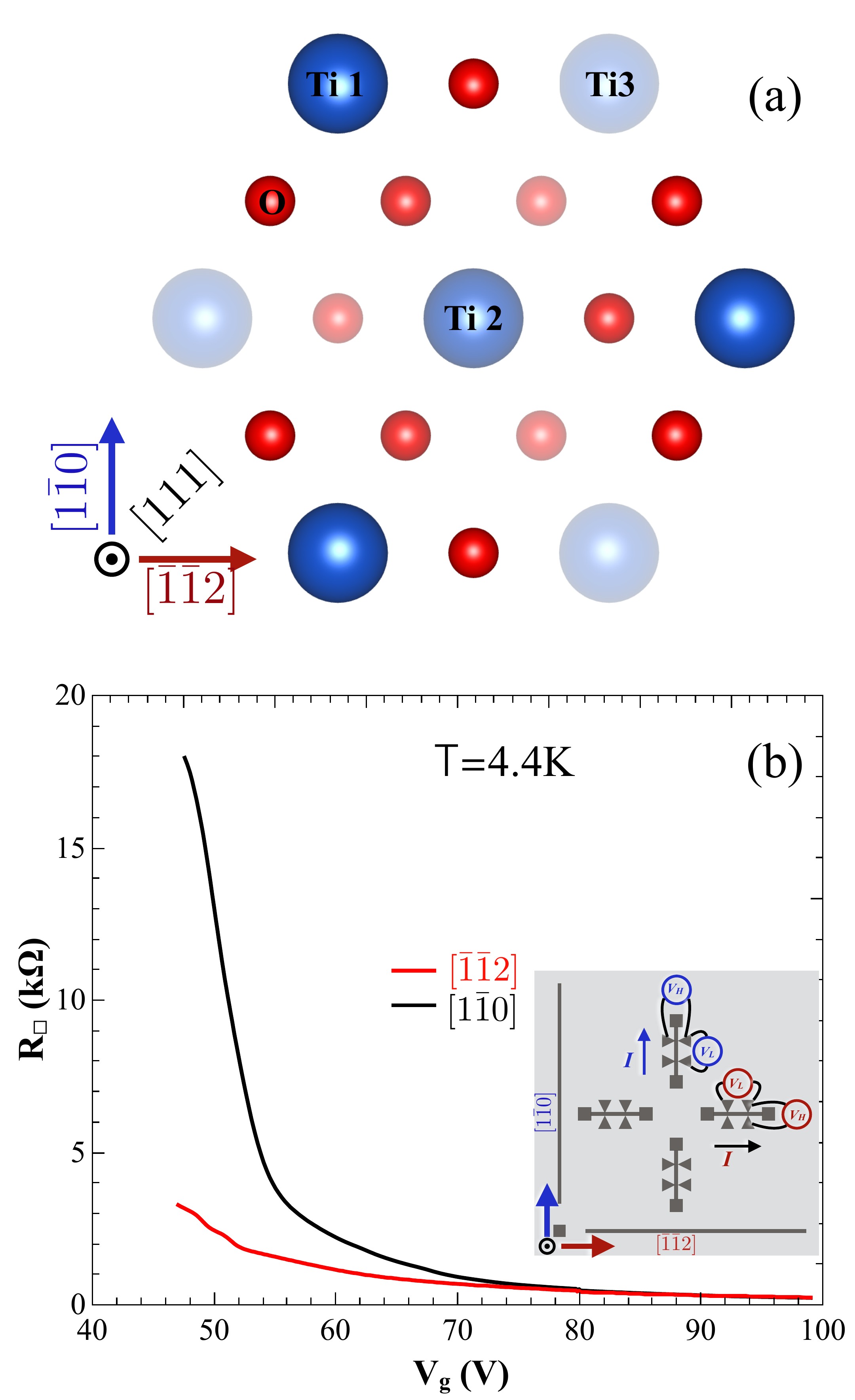}}
\vspace{-0.2cm}
\caption{a) Schematic representation of the first three monolayers at the LAO/STO interface with the  [$\bar{1}\bar{1}2$] and [$1\bar{1}0$] labeled. The red atoms represent the oxygen and blue atom titanium. The titanium atoms are further labeled Ti 1/2/3 to indicate their distance away from the interface, with Ti 1 being at the interface. (b) Averaged $R_\Box$ vs $V_g$ for the [$\bar{1}\bar{1}2$]/ [$1\bar{1}0$] in red/black measured simultaneously at 4.4 K. The inset shows a schematic of the device configuration on each measured LAO/STO sample. }
\vspace{-0.5cm}
\label{Gate}
\end{figure} 
Over more than a decade, the two-dimensional conducting gas that forms at the LAO/STO interface has been intensively studied because of the variety of properties that can be controlled through the application of an \textit{in-situ} electric field, including conductivity, superconductivity,\cite{reyren,dikin,mehta} ferromagnetism,\cite{dikin,mehta,brinkman,shalom2,Ariando} and the spin-orbit interaction.\cite{Gop,shalom}  Until recently, most of these studies focused on the (001) orientation of the LAO/STO heterostructures, while the (110) and (111) orientations have remained relatively unexplored.  The (111) orientation of the LAO/STO interface is especially interesting due to the hexagonal symmetry of the titanium atoms at the interface, shown schematically in Fig. \ref{Gate}(a).  This configuration has been likened to a strongly correlated analogue of graphene, and has been predicted to exhibit topological properties, unconventional superconductivity, as well as nematic phases.\cite{Doe,scheu,arun} Electric transport measurements have shown that the (111) LAO/STO interface exhibits many of the properties already seen in (001) LAO/STO, including a coexistence of superconductivity and magnetism.\cite{davis2,davis1,davis4,rout}  However, the feature that distinguishes the (111) interface from the (001) interface is the strong anisotropy with respect to surface crystal direction seen in almost all properties, including conductivity, Hall effect, superconductivity, quantum capacitance and longitudinal magnetoresistance.\cite{davis2,davis1,davis4}  By measuring the temperature dependence of the resistance and Hall coefficient, we show here that this anisotropy may be a signature of an electronic nematic transition that onsets at low temperatures, far from any known structural transitions of the system.

Four 100 $\mu$m by 600 $\mu$m Hall bar devices were fabricated on a substrate of (111) STO on which 20 monolayers of LAO were deposited epitaxially via pulsed laser deposition. The devices were oriented such that the lengths of two Hall bars lay along the [$\bar{1}\bar{1}2$] surface crystal direction and the lengths of the other two devices lay along the orthogonal [$1\bar{1}0$] crystal direction, as shown schematically in the inset of Fig. \ref{Gate}(b). Details of film synthesis and device fabrication can be found in earlier papers.\cite{davis2,davis1} To remind the reader of the striking anisotropy seen in these (111) LAO/STO heterostructures, Fig. \ref{Gate}(b) shows the longitudinal sheet resistance $R_\Box$ as a function of the applied back gate voltage $V_g$, measured along the two crystal directions at 4.4 K.  (As is seen in almost all LAO/STO structures, the resistance is a hysteretic function of $V_g$ due to the glassiness of the system at low temperatures, hence the traces in Fig. \ref{Gate}(b) are the averages of the up and down gate voltage sweeps, representing the long time-scale behavior of the system.)  At large positive gate voltages, the resistances along both directions are the same, but as $V_g$ is reduced, they begin to diverge, and by $V_g$= 45 V, the resistance along the [$1\bar{1}0$] direction is 5 times the resistance along the [$\bar{1}\bar{1}2$] direction.  In an earlier study,\cite{davis1} we showed that the number of oxygen vacancies in the sample influences the amount of anisotropy: the smaller the number of oxygen vacancies, the greater the anisotropy, and the higher the gate voltage at which the anisotropy turns on.  In comparison to samples in previous studies, the relatively large positive $V_g$ at which the anisotropy turns on, the low resistivity at large positive $V_g$, and the sharp increase in resistance with decreasing $V_g$ below $V_g\sim 70$ V seen in the present study indicates that these samples, in addition to having a small density of oxygen vacancies, also have a smaller amount of disorder, so that the effects that we observe arise primarily from the intrinsic band structure rather than defects.        

\begin{figure}[h!]
\center{\includegraphics[width=7cm]{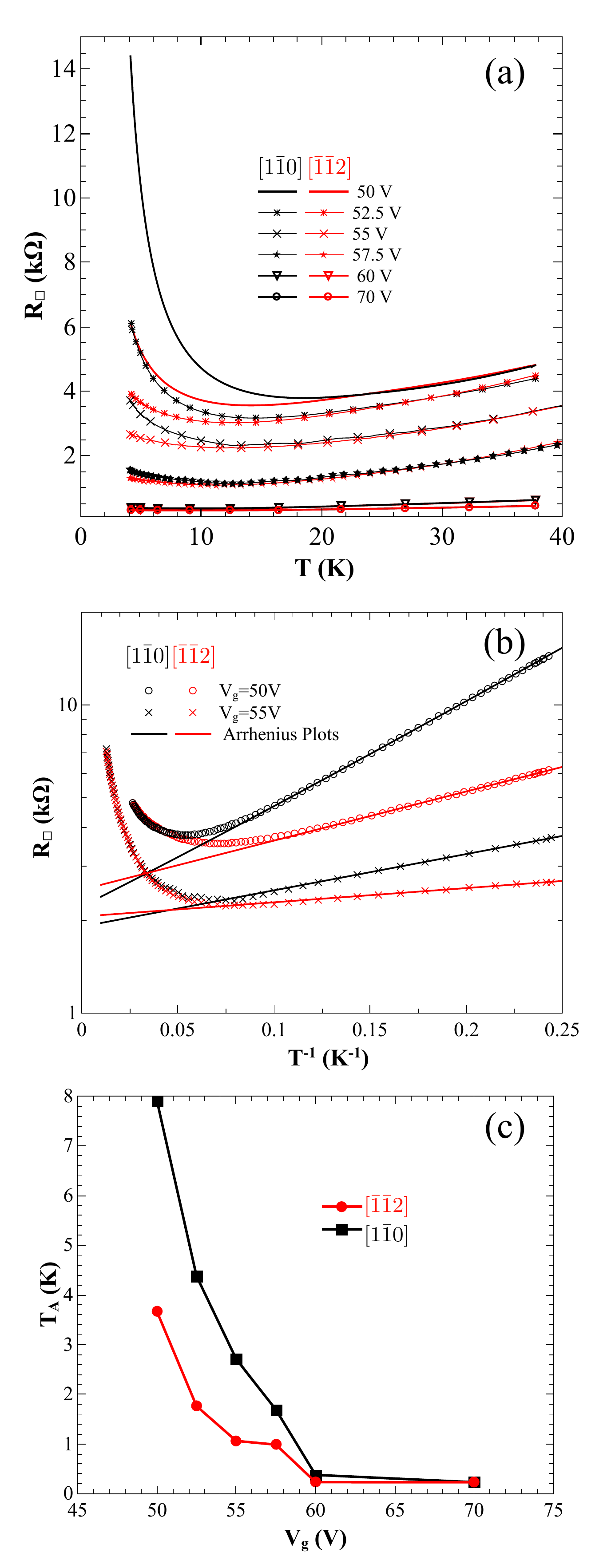}}
\caption{(a) $R_\Box$ vs $T$ for the [$\bar{1}\bar{1}2$]/ [$1\bar{1}0$] directions in red/black measured simultaneously at $V_g$=50 V. Measurable anisotropy can be seen below $\sim 22$ K.  (b) $R_\Box$ vs $1/T$ for $V_g$=50V and 55V superimposed with fits to activated behavior as described in the text. (c) Plots of the extracted activation temperatures $T_a$ vs $V_g$ for the [$\bar{1}\bar{1}2$]/ [$1\bar{1}0$] directions in red/black respectively. }
\vspace{-0.7cm}
\label{Longitudinal}
\end{figure}
STO is well-known to have a structural transition from a cubic to tetragonal crystal structure at around $\sim$105 K:\cite{Viana} previous studies have reported evidence of anisotropy associated with this transition in (001) LAO/STO.\cite{Frenkel,ma}  The direction of this anisotropy is random, as the system breaks up into tetragonal domains with randomly oriented $c$-axes.\cite{Frenkel}. There is evidence that the domain structure survives cycling to room temperature, but is still random from one device to the next.\cite{ma}  In contrast, the anisotropy we observe in the (111) LAO/STO devices is systematic across multiple cooldowns of the same sample and across tens of different samples:  the resistance measured along the [$1\bar{1}0$] is always larger than the resistance along the [$\bar{1}\bar{1}2$] direction at lower values of $V_g$, and this difference grows as $V_g$ is reduced.\cite{davis2,davis1}  In addition, as reported previously, the anisotropy in resistance is not observed at room temperature or at 77 K, ruling out the possibility that it is associated with the structural transition at $\sim$105 K. \cite{davis2}

In order to determine the temperature at which the anisotropy onsets, we have measured the resistance of the Hall bars along the two mutually perpendicular directions as a function of temperature, at different values of gate voltage in the range 50 V $\leq V_g \leq$ 70 V.  These data are shown in Fig. \ref{Longitudinal}(a).  At temperatures above $\sim$ 22 K, there is no difference in resistance between the two crystal directions.  As the temperature is reduced below 22 K, the resistances of the Hall bars along the two directions begin to differ, the difference growing with decreasing temperature.  As one might expect from Fig. \ref{Gate}(a), the anisotropy divergence is more pronounced as $V_g$ decreases.  From Fig. \ref{Longitudinal}(a), it appears that the temperature at which the anisotropy manifests itself changes with $V_g$; however, analysis at higher resolution (not shown) indicates that the temperature at which the anisotropy onsets is approximately the same ($\sim$ 22 K) for all the gate voltages shown.  Consequently, there appears to be a characteristic temperature, $T_{nem}$, well below the structural transition temperature of STO, that marks the onset of nematicity in this system.

Surprisingly, at temperatures $T < T_{nem}$, for $V_g<60$ V, the resistance shows an activated temperature dependence.  Figure \ref{Longitudinal}(b) shows the resistances for the two directions for $V_g$=50 V and $V_g$=55 V on a logarithmic scale as a function of $1/T$, along with fits to the function $R_\Box \sim e^{T_A/T}$ (only two gate voltages are shown for clarity).  As can be seen, the activated form describes the low temperature behavior of the resistance remarkably well (the fits are similar for the other gate voltages shown in Fig. \ref{Longitudinal}(a)).  Figure \ref{Longitudinal}(c) shows the resulting activation temperatures $T_A$ obtained from the fits as a function of $V_g$.  As expected, the activation temperatures for the two directions are different, but what this plot clearly shows is that the activated behavior for these devices turns on only below $V_g\sim60$ V.

\begin{figure}[!]
\center{\includegraphics[width=7cm]{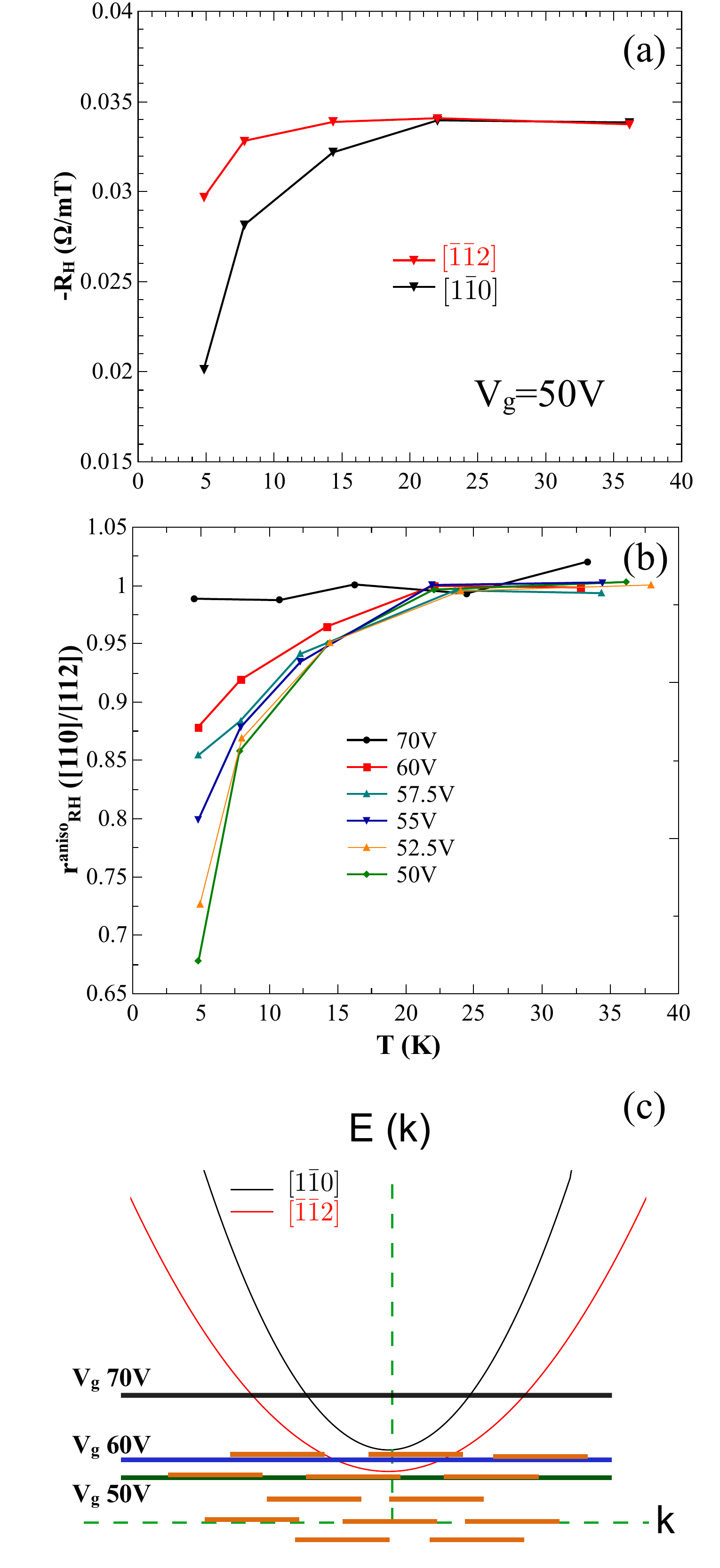}}
\caption{(a) The Hall coefficient $R_H$ vs. temperature $T$ for the [$\bar{1}\bar{1}2$] and [$1\bar{1}0$] directions in red and black respectively, measured simultaneously at $V_g$=50 V. (b) The ratio $r_H^{aniso} = R_{H_{[1\bar{1}0]}}/R_{H_{[\bar{1}\bar{1}2]}}$ of the Hall coefficients along the two crystal directions vs. $T$ for different $V_g$.   (c) Schematic representation of a system with anisotropic band edges. The solid parabolic bands represent the [$\bar{1}\bar{1}2$]/ [$1\bar{1}0$] directions in red and black respectively, the horizontal lines represent the Fermi energy at different values of $V_g$, and the orange dashed lines represent defect states at the interface.}
\vspace{-0.8cm}
\label{Hall}
\end{figure} 

Similar in-plane anisotropy has been observed in a variety of bulk materials such as Sr$_3$Ru$_2$O$_7$,\cite{Borzi} iron based superconductors like Ba(Fe$_{1-x}$Co$_x$)$_2$As$_2$,\cite{kim,kasa,yi} as well as two-dimensional electron gases, specifically GaAs/Al$_x$Ga$_{1-x}$As heterojunctions.\cite{cooper}  In the last example, the anisotropy arises from the emergence of a striped charge density wave in the $\nu =9/2$ state, and can exhibit a difference of resistance of up to a factor of 15.  In the case of iron based superconductors like Ba(Fe$_{1-x}$Co$_x$)$_2$As$_2$, this anisotropy is driven by a combination of a structural transition from tetragonal to orthorhombic crystal symmetry,\cite{kim} the onset of charge/orbital,\cite{yi} and spin order\cite{kasa}, giving rise to a robust nematic phase that is sensitive to both temperature and doping.  These materials show a more modest difference in in-plane resistivity, up to a factor of 2, but like the LAO/STO interface, play host to a wide variety of material phases such as superconductivity and magnetism. Aside from the anisotropy observed in $R_\Box$, another signature of nematicity observed in the iron based superconducting materials is a lifting of the degeneracies in carrier density and onsite binding energy of the Fe $d_{xz}$ and $d_{yz}$ electrons as measured by ARPES.\cite{yi}  While it is very difficult to measure an ARPES response from a buried LAO/STO interface, one can measure the Hall coefficient $R_H$ along both crystal directions to get an indication of the anisotropy in carrier density $n$ along both directions. Both LAO and STO are cubic at room temperature. and thus $R_H$ should be explicitly isotropic in the linear response regime.  Thus, any sign of anisotropy in $R_H$ that we observe would be further evidence of symmetry breaking and a possible nematic state. 

In our regime of temperature and gate voltage, the Hall response is linear out to $\pm100$ mT and the longitudinal magnetoresistance is flat to within $5$ $\Omega$; thus $R_H$ is simply the slope of the transverse resistance with perpendicular magnetic field.  Figure \ref{Hall}(a) shows the Hall coefficient $R_H$ as a function of temperature $T$ measured simultaneously along the [$1\bar{1}0$] and [$\bar{1}\bar{1}2$] directions at $V_g$ = 50 V.  Above $\sim$20 K, $R_H$ measured along both directions is the same, but below this temperature, $|R_H|$ measured along the [$1\bar{1}0$] direction decreases more rapidly with decreasing $T$ in comparison with $|R_H|$ measured along the [$\bar{1}\bar{1}2$] direction.  Similar behavior is observed for other values of $V_g \leq 60$ V.  In order to show the development of the anisotropy with decreasing $T$ for all values of $V_g$, we plot in Fig. \ref{Hall}(b) the ratio $r_H^{aniso} = R_{H_{[1\bar{1}0]}}/R_{H_{[\bar{1}\bar{1}2]}}$ of the Hall coefficients along the two crystal directions as a function of $T$, for different values of $V_g$.  From this plot, we confirm two observations made from the anisotropic resistance measurements:  first, the anisotropy develops only for $V_g \leq 70$ V (the $V_g=70$ V data for both $R_\Box$ and $R_H$ show no anisotropy); and second, for $V_g < 70$ V, the anisotropy develops only below a temperature $T_{nem} \sim 22$ K, the same temperature found for the longitudinal resistance measurements.  Thus, the anisotropic behavior seen in $R_\Box$ and $R_H$ are two manifestations of the same nematic transition.

While the arguments given earlier rule out the structural at $\sim$105 K as the source of the anisotropy, there have been reports of anomalies in the dielectric constant of STO at lower temperatures that might be related to our observations.\cite{Viana}  In particular, a small peak has been reported in the dissipative component of the dielectric constant in the temperature range of 20-30 K that might be linked with the well-known quantum paraelectric transition at 37 K.\cite{Viana}  At present, it is not clear how these features in the dielectric response of the STO substrate would lead to an anisotropy in almost all measured characteristics.  Nevertheless, our experimental observations suggest certain necessary characteristics of the band structure of (111) LAO/STO that any mechanism must give rise to in order to explain the anisotropy.  When current is injected along the [$\bar{1}\bar{1}2$] direction, it is carried by carriers in a band whose band edge is lower in comparison to the corresponding band for the [$1\bar{1}0$] direction.  This is shown schematically in Fig. \ref{Hall}(c), where the [$1\bar{1}0$] band is in black and the [$\bar{1}\bar{1}2$] is in red.  Consider the case when scattering between these bands is suppressed for some reason, so that a carrier injected into one band cannot scatter into the other.  For large gate voltages ($V_g \geq 70$ V in the present case), the Fermi level is such that bands along both directions are substantially occupied, so that no anisotropy would be observed.  As $V_g$ is decreased, the Fermi level first drops below the band edge corresponding to the [$1\bar{1}0$] direction, then the band edge corresponding to the [$\bar{1}\bar{1}2$] direction.  When the Fermi level drops below a band edge, the only way of getting carriers into the band is by thermal excitation from localized levels in the band gap (shown schematically in Fig. \ref{Hall}(c)), leading to the activated behavior, with the activation temperature $T_A$ along the [$1\bar{1}0$] direction being larger than $T_A$ along the [$\bar{1}\bar{1}2$] direction, as we observe.           

In order to sustain the nematicity we observe, scattering between the bands contributing to transport along the two orthogonal directions must be suppressed, otherwise any observable differences between the transport characteristics of the two bands would be washed out.  Even with the relatively clean samples here, one expects the degree of disorder, and hence the scattering of carriers to be large, so it is hard to imagine that a strong suppression of inter-band scattering would occur.  However, in systems with strong spin-orbit scattering, it is known that carrier bands may become spin-polarized so that spin-momentum locking may occur.\cite{hsieh,zeng}  In this case, a carrier may scatter to another band only if its spin is also flipped; such scattering would then be suppressed in the absence of spin-flip scattering mechanisms.  It is well-known that strong spin-orbit interactions are present at the LAO/STO interface,\cite{Gop,shalom,syro2,banerjee} and are a likely source of the splitting of the band edges of the otherwise degenerate bands. Furthermore large spin-orbit coupling is predected to stabilize nematic behavior in the sample, even in the presence of large disorder.\cite{arun} Consequently, we believe the model presented here is the simplest explanation of the origin of the anisotropic behavior we observe, although more exotic mechanisms such as the nucleation of a charge density wave cannot be ruled out at present. While computations of the band structure of (001) LAO/STO interface structures\cite{banerjee,khalsa} as well (111) LAO/STO superlattices\cite{Doe} taking into account spin-orbit interactions and structural transition have been made, it would be interesting to perform similar calculations on simple (111) LAO/STO interfaces to understand the role of spin-orbit interactions coupled with structural changes on the band structure.

\section*{Methods }The 20 monolayer (ML) (111) LAO/STO interface samples reported on in this study were prepared by pulsed laser deposition using a KrF laser ($\lambda$ = 248 nm).  A single crystal LAO target was used for deposition, with a laser repetition of 1 Hz, laser fluence at 1.8 J/cm$^2$, growth temperature at 650 C, and oxygen pressure at 0.1 mTorr.  The deposition was monitored via \textit{in situ} reflection high energy electron diffraction (RHEED). Hall bars were then fabricated using photolithography to define an etch mask.  Argon ion milling was then used to etch the unmasked areas to the bare STO, leaving the LAO on top of the Hall bar devices.  A final lithography step was used to deposit a Au film on the electrical contacts to enable visual location of the contacts for wire-bonding as well as control structures on the bare STO to ensure the bare STO did not become conducting.  As detailed in our previous work,\cite{davis2} AFM characterization of the atomic steps showed that occur at 45$^{\circ}$ angles to all four hall bars, thus the anisotropy observed is not due to crystal terrace effects.

The devices were measured using a homemade cryostat and conventional lock-in techniques.  The lock-in techniques utilized homemade low noise current sources and amplifiers and to measure Hall leads or longitudinal leads.  Temperature was continuously monitored via a silicon diode thermometer and homemade temperature bridge.  As described in our previous work, \cite{davis2} the average of the up and down sweep of $R_\Box$ vs $V_g$ is taken as the long time scale behavior of the system.

\begin{acknowledgments}
Work at Northwestern was funded through a grant from the U.S. Department of Energy through Grant No. DE-FG02-06ER46346.	Work at NUS was supported by the MOE Tier 1 (Grant No. R-144-000-364-112 and R-144-000-346-112) and Singapore National Research Foundation (NRF) under the Competitive Research Programs (CRP Award Nos. NRF-CRP8-2011-06, NRF-CRP10-2012-02, and NRF-CRP15-2015-01).

\end{acknowledgments}

\end{document}